\documentclass{iopjournal}

\usepackage{float}

\usepackage{ragged2e}
\justifying

\begin{document}

\articletype{Article type} 

\title{Functional Interface Blocks for Neuromorphic Hardware: A Junction-Centered Framework}

\author{Wellington Avelino$^{1,*}$\orcid{0000-0002-9273-9827}, Yann Beillard$^3$\orcid{0000-0003-0311-8840}, Fabien Allibart$^4$\orcid{0000-0002-9591-220X}, Dominique Drouin$^3$\orcid{0000-0003-2156-967X} and Gilberto Medeiros-Ribeiro$^2$\orcid{0000-0001-5309-2488}}

\affil{$^1$Department of Electrical Engineering, University of Minas Gerais, Belo Horizonte, Brazil}

\affil{$^2$Computer Science Department, University of Minas Gerais, Belo Horizonte, Brazil}

\affil{$^3$Department of Electrical and Computer Engineering, Université de Sherbrooke, Sherbrooke, Canada}

\affil{$^4$Nanotechnology Department, 3iT - Interdisciplinary Institute for Technological Innovation , Sherbrooke, Canada}

\affil{$^*$Author to whom any correspondence should be addressed.}

\email{wavelino@ufmg.br / wellingtonavelino@ymail.com}


\begin{abstract}
\justifying
Heterogeneous neuromorphic hardware integrates devices with dissimilar electrical characteristics and dynamics, making functional compatibility at their interconnections a primary design challenge. Direct coupling alone is insufficient to ensure correct operation, because the load-line conditions established at each junction determine the effective operating regime. Here, we propose a junction-centered interface framework in which inter-device connections are described through assigned drive/sense roles and organized into canonical functional interface blocks. As a concrete hardware realization, a second-generation current conveyor (CCII)-based implementation is then adopted as a composite realization of these interface primitives. The framework is validated experimentally in a Pavlovian-conditioning demonstrator combining a memristive synapse with a unijunction-transistor (UJT) post-neuron. By linking local junction conditions to reusable interface functions, the proposed methodology provides a systematic basis for the design and analysis of heterogeneous neuromorphic systems.
\end{abstract}

 \noindent\textbf{Preprint notice.}
 This manuscript is a preprint version and has not yet undergone peer review. It has been prepared for possible submission to \textit{Neuromorphic Computing and Engineering.}

\section{Introduction}
\justifying
Neuromorphic hardware has emerged as a promising route toward computation by embedding information processing directly in the physical operation of electronic devices and circuits. In this framework, voltage, current, and internal state variables are not merely auxiliary quantities, but degrees of freedom that participate in computation \cite{Mead2002NeuromorphicElectronicSystems}. This perspective has motivated the exploration of a broad range of neuromorphic platforms, including analog and mixed-signal CMOS circuits \cite{Indiveri2011SiliconNeuronCircuits}, memristive devices \cite{Adhikari2012MemristorBridge}, thyristor-based neurons \cite{Rozenberg2019UltraCompactLIF}, Mott-transition devices \cite{Pickett2013ScalableNeuristor}, spintronic elements \cite{Grollier2020NeuromorphicSpintronics}, digital spiking realizations on reconfigurable hardware \cite{Sowmya2023NeuromorphicFPGA}, and active transmission lines \cite{Brown2024AxonLike}. However, although neuromorphic performance depends on device physics and circuit operation \cite{Markovic2020physics}, the electrical interfaces that couple neurons and synapses are often treated as secondary integration constraints. In real hardware, neurons and synapses interact through electrical terminals, making impedance and loading conditions central to the computation itself.
Driving artificial neurons implemented in hardware is not equivalent to applying an abstract input to a mathematical model. A hardware neuron is a nonlinear dynamical circuit whose state may be encoded in membrane voltage \cite{Hodgkin1952propagation}, stored charge \cite{Rozenberg2019UltraCompactLIF}, internal conductance \cite{Mahowald1991silicon}, leakage current \cite{Indiveri2011SiliconNeuronCircuits}, lattice temperature \cite{Del2019subthreshold}, or other device-dependent variables. As a result, the input signal delivered by a synapse interacts with finite and state-dependent impedance, and time constants associated with integration, thresholding, leakage, and reset \cite{Chicca2014NeuromorphicCircuits}. Likewise, the output spike must drive subsequent synaptic loads without distortion or unintended modification of their internal states. Thus, neuron driving is not merely an implementation detail, but a fundamental interface problem for preserving computation across coupled neuromorphic blocks, requiring impedance compatibility and control of spike amplitude, duration, and polarity at the synaptic terminals. This challenge becomes even more critical as neuromorphic platforms combine devices governed by distinct switching mechanisms and electrical interface conditions. Mismatches in voltage range, current level, impedance, pulse requirements, and internal state dynamics can hinder reliable communication between elements such as current-controlled negative differential resistance (CC-NDR) devices, voltage-controlled negative differential resistance (VC-NDR) devices, memristors, phase-transition devices, spintronic elements, and CMOS neuron circuits \cite{Markovic2020physics}.
A reciprocal constraint appears when neuronal spikes drive memristive synapses: the spike is not only a communication signal, but also a programming or readout stimulus applied to a memory device.  Memristive synapses impose distinct drive requirements during inference and training: their conductance must be sensed non-invasively during readout but programmed through controlled electrical stress during weight update. In filamentary devices, SET typically requires current compliance to prevent uncontrolled filament growth, while RESET depends on voltage polarity, amplitude, and pulse duration \cite{Yang2008memristive}. In non-filamentary switching devices, conductance updates arise from spatially distributed ionic or electronic processes, making them sensitive to waveform and programming history \cite{Wang2015TaTaOxTiO2SynapticDevice}. In both cases, operation is further limited by stochastic switching, cycle-to-cycle and device-to-device variability, endurance and retention constraints, and nonlinear or asymmetric conductance updates.
These device-level constraints become system-level interface constraints when memristive synapses are coupled to neuron circuits, which often operate at different voltage and current scales. In such cases, the realized operating point is determined by the intersection of their load characteristics, so mismatched load lines may distort membrane integration, overload neuron inputs, or disturb stored weights. Avoiding these effects requires voltage and current regulation, compliance control, impedance matching, and read/write isolation \cite{Garg2024AnalogLIF}. Although hybrid neuromorphic systems can be achieved through technology-specific co-design and peripheral circuitry \cite{Huang2024MemristorAccelerators,Koo2025CommercialEmergingMemory,Li2023MemristiveNeuronSynapse}, such solutions are typically platform-specific, motivating a more general and physically grounded framework for neuromorphic interfaces.
To address these limitations, this work introduces the functional interface block (FIB) as a hardware-level abstraction for sensing, conditioning, and delivering signals between neuromorphic elements. By separating the functional coupling between neuromorphic elements from the electrical conditions required at their interface, the FIB allows voltage-driven, current-driven, mixed-mode, and state-dependent devices to be described within a common framework. To experimentally ground this framework, we implement the FIB using a second-generation current conveyor (CCII) \cite{SedraSmith1970CurrentConveyor}, a classical circuit topology well suited to define controlled voltage–current sensing and driving relations \cite{Lecerf2014silicon}. The CCII is therefore used not as a unique interface solution, but as an accessible reference platform to evaluate FIB operation in a spiking neural network performing associative learning. This implementation demonstrates reliable signal mediation and state-dependent interaction within a defined operating range, while providing a basis to assess non-idealities, operating limits, and scalability constraints in heterogeneous neuromorphic systems.
The main contributions of this work are twofold. First, we propose a device-agnostic FIB framework that integrates a junction-centered methodology for describing, placing, and specifying the sensing, conditioning, and driving functions required to couple heterogeneous neuromorphic elements, including memristive synapses and NDR-based neurons. Second, we implement the framework using a CCII-based reference hardware platform, enabling experimental validation and systematic evaluation of operating limits, non-idealities, and scalability constraints. The remainder of the paper is organized as follows: Section 2 motivates interface decoupling in heterogeneous neuromorphic systems; Sections 3 and 4 formalize FIB placement and taxonomy; Sections 5 and 6 present and validate the CCII-based implementation; and Section 7 concludes the paper.

\section{Why must heterogeneous neuromorphic interactions be decoupled}
\begin{figure}
 \centering
        \includegraphics[width=0.80\textwidth]{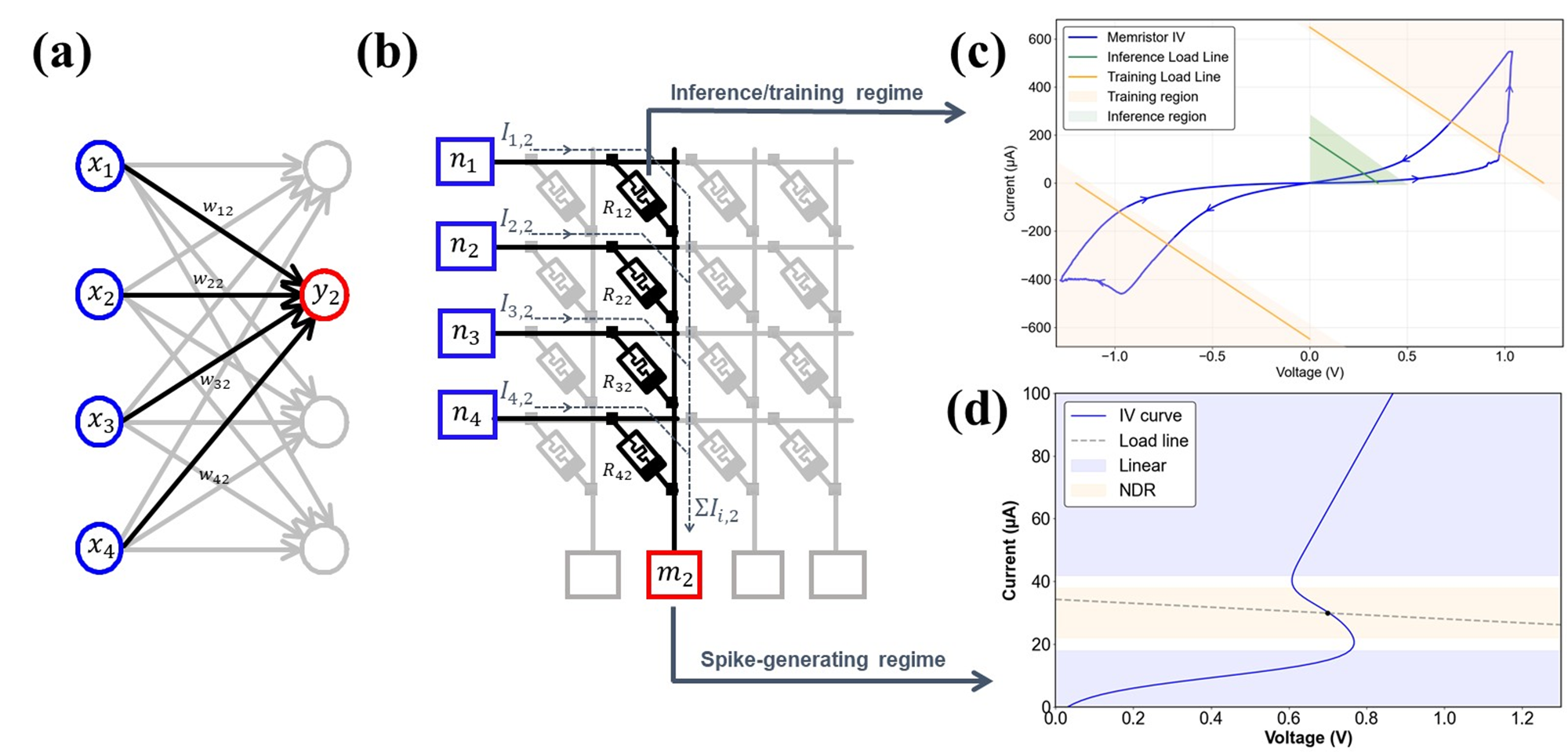}
 \caption{
 \justifying
 From neural-network abstraction to conditioned physical interaction enabled by the FIB. (a) Neural-network representation, where computation is described at the functional level (neurons and weighted synapses). (b) Physical realization of the same network in a heterogeneous crossbar-like hardware substrate, where memristive synapses ($R_{12}$-$R_{42}$) connect input neurons ($n_1$-$n_4$) to a postsynaptic node, resulting a summed signal $i2$ that drives the spiking neuron $m_2$. (c) Load-characteristic mismatch illustrated under standalone biasing (dashed lines) for a memristive synapse and (d) switching neuron component, which is composed by a NDR-based element (or hysteresis): each device typically requires different external bias conditions to reach its intended operating regime; under direct interconnection, a single shared constraint is enforced, which shows that synaptic and neuronal devices generally exhibit different electrical characteristics and preferred operating regions; when directly interconnected, they are forced to share a common electrical constraint, which can displace one or both devices from their desired regime. This motivates the use of impedance-conditioned interfacing, in which the effective input/output conditions are matched to the electrical variable being sensed or enforced.}
\label{fig1}
\end{figure}

\justifying
Neural-network diagrams describe computation in terms of signals exchanged between neurons and synapses, but they do not specify the electrical constraints required for physical implementation. This gap is illustrated in Fig.\ref{fig1}(a,b), where the same network is shown first as a graphical abstraction and then as an electrical interconnection of physical devices. Once instantiated in hardware, the same network becomes an interconnection of dissimilar devices, each operating under specific bias, impedance, and frequency conditions. The design problem therefore shifts from defining which elements are connected to determining how they electrically interact at their junctions.

\noindent In physical implementations, device behavior is determined by the intersection between intrinsic device characteristics, the associated parasitics, and the external electrical circuitry imposed at the interconnection. The realized operating point is thus defined not by device physics alone, but by the load line and impedance environment at the network junction \cite{Sedra2011MicroelectronicCircuits}. Interface design therefore becomes central, since the surrounding circuitry effectively selects which region of the device characteristic is accessed during operation. Consequently, in heterogeneous neuromorphic systems, the junction itself is a key design challenge.
This issue becomes especially salient when memristive synapses and spiking neurons are directly coupled. Memristors typically require different voltage and current regimes for read and programming operations: inference is usually performed under low-amplitude, non-disturbing bias conditions, often quasi-unipolar, whereas training requires larger state-changing excursions that may be unipolar or bipolar depending on the switching mechanism and device technology, exercised under current compliance, constant current, or constant voltage \cite{Wang2020ResistiveSwitchingMaterials}, as illustrated in Fig.\ref{fig1}(c). By contrast, spiking-neuron implementations based on switching devices rely on strongly nonlinear characteristics such as hysteresis or NDR \cite{Sarkar2022OrganicSpikingNeuron, Kim2024BiomimeticSpikingNeuron} that can be voltage or current controlled, whose key role is to provide thresholded regenerative switching and a distinct return/reset trajectory. These properties are fundamental for excitability and pulse generation, since they enable abrupt state transitions and, together with the surrounding load and capacitance, support relaxation-like spiking dynamics, as illustrated in Fig.\ref{fig1}(d).
When such devices are directly interconnected, both must satisfy a shared electrical constraint (source and input impedances, voltage or current drive, or sense), even though their optimum operating regions are generally different. As a result, one or both elements may be not properly biased or driven, degrading synaptic read/program separation, suppressing the neuron’s switching dynamics, or preventing correct learning behavior \cite{Sedra2011MicroelectronicCircuits}. The key requirement is therefore to couple them electrically while keeping them at their optimum operating condition and thus preserving the transfer of information across their junctions. In heterogeneous neuromorphic hardware, this information remains encoded in electrical variables such as voltage, current, or spike waveforms, and must be conveyed in a form that is compatible with the operating regime and impedance requirements of the receiving device. The interface must therefore fulfill a dual role: it must prevent one device from forcing the other out of its intended operating region, while still sensing the relevant electrical state on one side and driving a compatible variable on the other under appropriate impedance conditions. 
Although the present discussion is motivated by core-domain synapse–neuron coupling, the same interface principle applies more broadly at the network boundaries, where external variables must be transduced into neural-compatible electrical signals and post-neuron activity must be conditioned for delivery to subsequent layers or event-routing circuitry, like active transmission lines. These requirements are not static, since the same junction may need to sustain different constraints in inference and training, often under time-multiplexed operation.  Essentially, rather than treating interfacing as a local circuit fix, we identify the network junctions at which Functional Interface Blocks (FIBs) must be inserted and define, for each operating mode, the Drive/Sense role of each port together with the impedance conditions required to preserve functional information transfer while electrically decoupling dissimilar devices.

\section{Junction-Centered Placement of Functional Interface Blocks}
\begin{figure}
    \centering
        \includegraphics[width=0.95\textwidth]{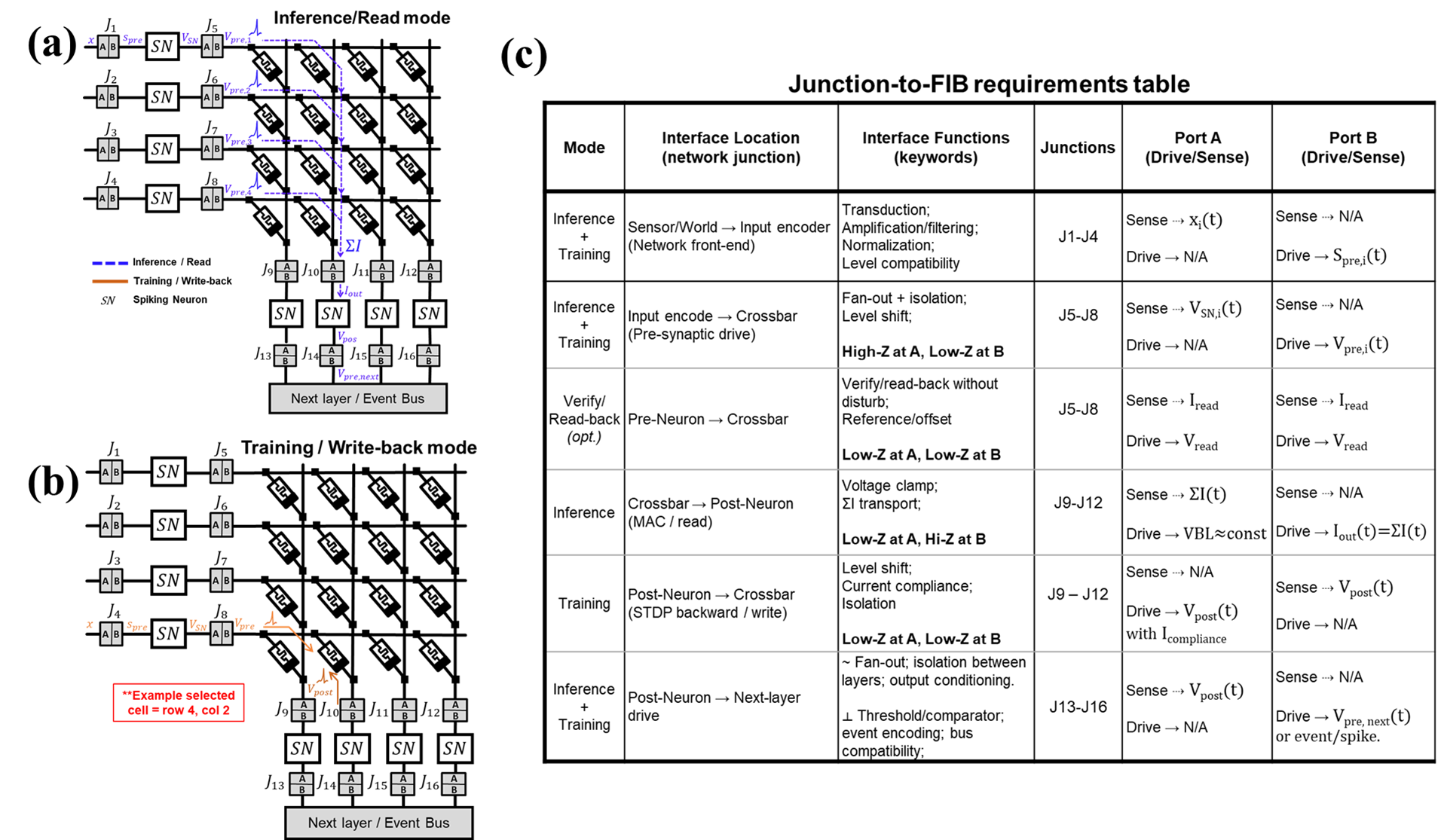}
    \caption{
        \justifying
Junction-to-FIB placement guide for a crossbar-based SNN. (a) Inference/read mode: External input variables $x_{i}(t)$ are conditioned at the front-end junctions $J_1$-$J_4$, passed through the pre-neuron stage, and converted into pre-synaptic row-driving signals applied at $J_4$-$J_8$. The crossbar performs the weighted summation, and the resulting column current is accumulated as I(t) at the column junctions $J_9$-$J_{12}$, where it is conveyed to the post-neuron stage. (b) Training/write-back mode: The same junctions $J_5$-$J_{12}$ are reused in time-multiplexed operation to apply programming stimuli to selected cells (example shown row 4, column 2), enabling backward/STDP-style updates while enforcing write isolation and current compliance. (c) unction-to-FIB requirements table: Each network junction $J_1$-$J_{16}$ is mapped to the required Drive/Sense roles at Port A and Port B and to the corresponding interface functions. Junctions $J_5$-$J_{12}$ form the core-domain interface blocks, i.e., the load-line- and impedance-critical read/write path between crossbar and neuron. Junctions $J_1$-$J_4$ and $J_{13}$-$J_{16}$ form the architectural/peripheral interface blocks, which respectively (i) transduce and encode external variables into electrical signals compatible with the network input, and (ii) condition post-neuron activity for delivery either to a subsequent crossbar layer or to an event bus.}
\label{fig2}
\end{figure}

\justifying
Connections between elements of heterogeneous neuromorphic circuitry constitute the central design challenge, because interface requirements are determined not only by device-level characteristics but also by the mode of operation. The first step, therefore, is to identify where FIBs must be inserted into the network. In this work, a junction is defined as any location at which signal type (voltage or current), impedance, and/or mode of operation (training/inference) changes. Figures \ref{fig2}(a) and \ref{fig2}(b) provide this junction map for inference and training operation, respectively. Also shown is the signal path from external input variables, through pre-neuron encoding and crossbar interaction, to post-neuron output and subsequent layer or event-bus delivery. The training diagram in Fig.\ref{fig2}(b) further highlights that the same physical network must also support as programming signals applied back to selected cells (write-back) by post-neurons \cite{ZamarrenoRamos2011STDP}, illustrating that junction placement must account not only for forward information flow, but also for mode-dependent (training or inference) reuse at the same physical locations.
The junction map of Fig.\ref{fig2} reveals that these locations fall into two distinct interface domains, which can include at most 4N junctions, where N is the number of bit/word lines. Junctions $J_1$–$J_4$ and $J_{13}$–$J_{16}$ belong to the peripheral domain, where signals respectively enter and leave the network core, whereas the core-domain junctions $J_5$–$J_{12}$ interface the  Spiking neurons to the Synaptic weights, where the computation process takes place. This distinction is essential because not all interfaces solve the same problem: peripheral interfaces primarily ensure compatibility between external or inter-layer variables (or simply, sensors and connections to the external world) and the network representation, whereas core-domain interfaces must directly mediate electrical compatibility between heterogeneous devices within the computational core. The table on Fig.\ref{fig2}(c) describes the mapping of each junction to its corresponding interface location and required function.
 
\noindent The core-domain interfaces occupy junctions $J_{5}$–$J_{12}$, where electrically encoded information is exchanged between the pre-neuron stage, the crossbar, and the post-neuron stage under the most load-line- and impedance-sensitive conditions. In this region, the role of the interface is to ensure that signals are transferred between stages without disturbing the voltage, current, and impedance conditions required by each device. Junctions $J_5$–$J_8$ support the word-line excitation path and are therefore associated with functions such as fan-out, isolation, level shifting, and the delivery of pre-synaptic drive signals to the addressed rows. Junctions $J_9$–$J_{12}$ define the bit-line read/write path, where the accumulated column response must be conditioned for delivery to the post-neuron during inference and, when required, where write-back stimuli must be applied during training. As summarized in Fig.\ref{fig2}(c), these bit-line interfaces may therefore require functions such as voltage clamping, current transport, current compliance, and write isolation. Taken together, junctions $J_5$–$J_{12}$ form the impedance-critical interaction path of the network core, since they directly determine the effective operating conditions seen by both the synaptic and neuronal devices.
A key feature of the proposed methodology is that interface requirements are inherently mode dependent. As illustrated in Fig.\ref{fig2}(a,b), the same physical junction may operate under distinct electrical constraints during inference/read, optional verify/read-back, and training/write-back operation. Consequently, FIB placement cannot be defined independently of mode assignment: identifying where an interface is needed also requires specifying how its function changes over the operating cycle. This time-multiplexed reuse is particularly important in compact neuromorphic hardware, where the same junction must support both non-disturbing read conditions and state-changing write conditions without compromising compatibility, isolation, or controllability. The placement problem is therefore not only spatial, but also operational, since each junction must be interpreted in terms of the sequence of electrical roles it must sustain across modes”
These requirements are consolidated in the junction-to-FIB table of Fig.\ref{fig2}(c), in which each junction is mapped to physical ports (Port A and Port B), together with the corresponding mode-dependent Drive/Sense assignments and interface functions. This representation makes clear that interface specification precedes circuit selection: once the junction, operating mode, and port-level roles are identified, the required behavior of the interface can be inferred systematically. The resulting design flow is therefore not to choose a circuit first, but to proceed in the opposite direction: identify the junction, determine its role across modes, assign what each port must sense and drive, and only then derive the interface functions required to preserve compatibility and information transfer. Once made explicit in this form, these junction-level requirements can be abstracted into a compact set of canonical interface behaviors, which motivates the taxonomy introduced in the next section.

\section{Canonical Taxonomy of Functional Interface Blocks}
\begin{figure}[H]
    \centering
        \includegraphics[width=0.55\textwidth]{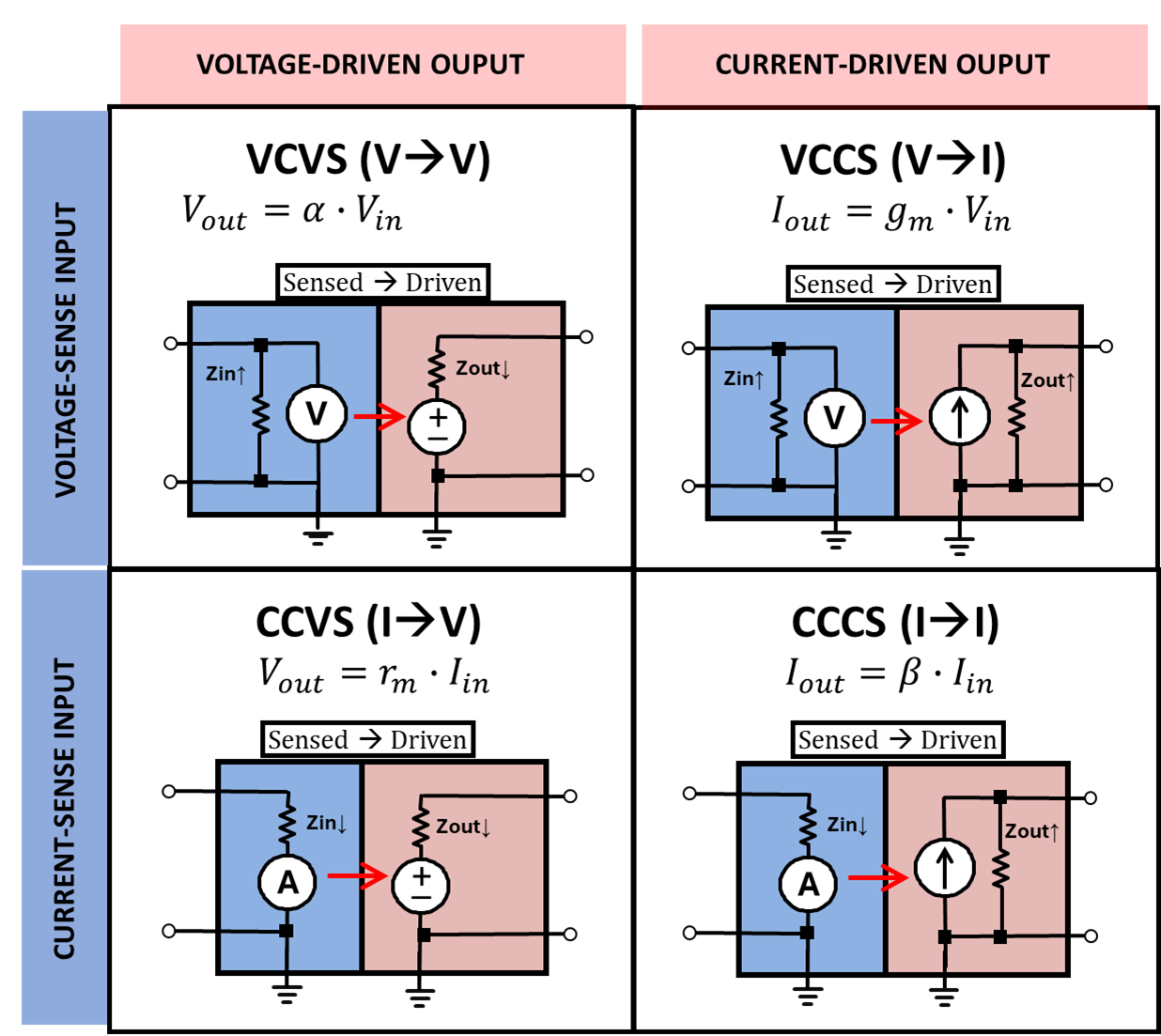}
    \caption{
        \justifying
Taxonomy of core FIB primitives. The interface blocks are organized by the electrical variable sensed at the input (rows) and the variable driven at the output (columns). Each cell depicts the corresponding primitive—VCVS, VCCS, CCVS, and CCCS—highlighting the key impedance requirements that enable functional decoupling: high input impedance for voltage sensing and low input impedance for current sensing, combined with low output impedance for voltage enforcement and high output impedance for current enforcement.}
\label{fig3}
\end{figure}

\justifying
Fig.\ref{fig3} introduces a canonical taxonomy of FIBs, organized according to the electrical variable sensed at the input and the variable driven at the output.

\noindent The taxonomy adopted here follows the canonical controlled-source classes of circuit theory \cite{AlexanderSadiku2021Fundamentals}. It is defined along two orthogonal axes: electrical variable sensed at the input side of the interface, which may be either voltage or current and the electrical variable driven at the output side, which may likewise be either voltage or current. Their combination yields four canonical classes: voltage-controlled voltage source (VCVS), voltage-controlled current source (VCCS), current-controlled voltage source (CCVS), and current-controlled current source (CCCS). In this formulation, the taxonomy does not yet refer to specific circuit implementation, but rather to the dominant transduction performed by the interface between its sensed and driven sides. It thus provides a compact functional description of what the interface must do before addressing how that behavior is physically realized.
Beyond the sensed-to-driven variable mapping, each canonical class also carries an implicit impedance interpretation at its input and output sides. Voltage sensing is naturally associated with a high-impedance input, so that the interface can observe the local electrical state without significantly perturbing it, whereas current sensing is associated with a low-impedance input that allows current to be captured under controlled potential conditions. Conversely, voltage driving implies a low-impedance output capable of enforcing a target potential at the receiving side, whereas current driving is associated with a high-impedance output that delivers current while allowing the terminal voltage to adjust according to the load. In this way, the taxonomy links variable compatibility to impedance conditioning: the interface is characterized not only by what electrical quantity it senses and drives, but also by the boundary conditions it presents to the connected devices.
These four classes should be understood as elementary interface primitives rather than complete circuit implementations. A given junction may require only one such primitive, or a practical interface may combine several of them within a composite block. In some cases, the same primitive may also be reused across different operating modes. The purpose of the taxonomy is therefore to classify the dominant input-output relation required at the interface, rather than to enumerate implementation details. Fig.\ref{fig3} defines this compact functional design space, while leaving room for practical realizations to differ in internal structure, complexity, and mode-dependent use.

\section{CCII-Based Realization of the Functional Interface Block}
\begin{figure}[H]
    \centering
        \includegraphics[width=0.90\textwidth]{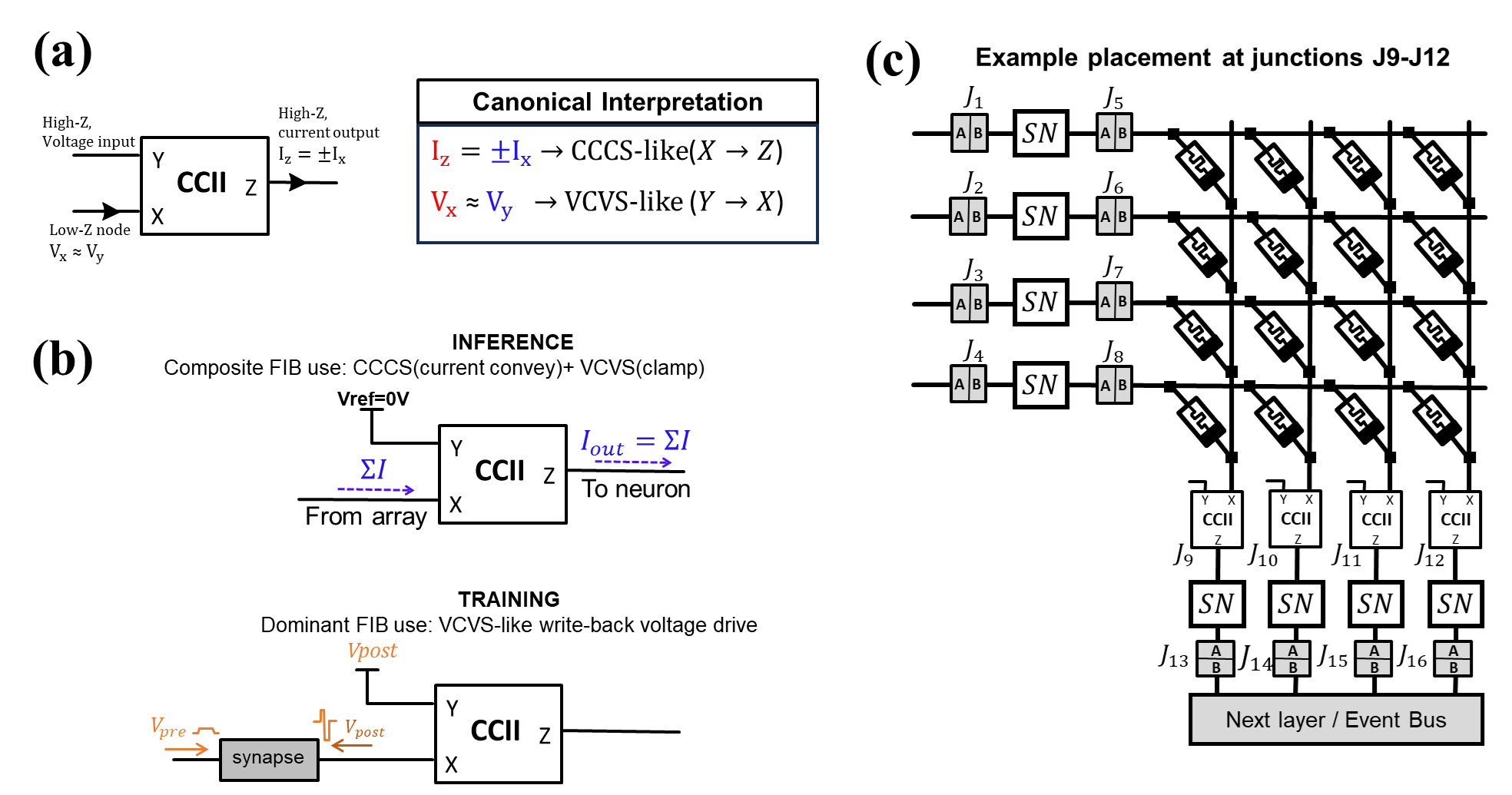}
    \caption{
        \justifying
CCII-based FIB as a practical realization of the interface taxonomy. (a) Ideal second-generation current conveyor and its canonical interpretation: terminal Y is a high-impedance voltage input, terminal X is a low-impedance node with $V_x$ $\approx$ $V_y$, and terminal Z is a high-impedance current output with $I_z$ = $\pm$$I_x$, corresponding to a composite VCVS-like (Y→X) and CCCS-like (X→Z) behavior. (b) Mode-dependent use of the selected CCII-based FIB. In inference, the block provides column-node clamping and conveys the accumulated array current $\sigma$I to the neuron ($I_{out}$=$\sigma$I). In training, the same block is reused in time-multiplexed form to deliver the post-neuron write-back voltage $V_{post}(t)$ to the array. (c) Example placement of CCII-based FIBs at the column junctions $J_9$–$J_{12}$, where they interface the crossbar with downstream neuron and event-bus circuitry.}
\label{fig4}
\end{figure}

\justifying
Having introduced a canonical taxonomy of FIBs, a practical realization of the required interface behavior is the next step. The FIB selected in this work is based on the second-generation current conveyor (CCII), \cite{Musil2009CurrentConveyor} interpreted as a composite implementation of the canonical primitives of Fig.\ref{fig3}, specifically a VCVS and CCCS, building on prior CCII-based neuron–memristor embodiments reported in the literature \cite{Lecerf2014silicon}. CCII naturally combines voltage transfer and current conveyance within a single structure, matching the mode-dependent requirements identified at the relevant crossbar-neuron junctions. Fig.\ref{fig4} makes this link explicit by relating the ideal CCII behavior to both the taxonomy of Fig.\ref{fig3} and the placement framework of Fig.\ref{fig2}.

\noindent Terminal Y behaves as a high-impedance voltage input, terminal X as a low-impedance node whose voltage follows $V_y$ ($V_x$ $\approx$ $V_y$), and terminal Z as a high-impedance current output that conveys the current entering $I_x$ ($I_z$ = $\pm$$I_x$). In the language of Fig.\ref{fig3}, these relations correspond to a composite interface behavior: a VCVS-like voltage transfer from Y to X, combined with a CCCS-like current transfer from X to Z. This interpretation is particularly relevant for the present application because it allows the same block to impose a controlled voltage condition at the interaction node while simultaneously extracting and transporting the associated current, thereby preserving functional information flow without sacrificing electrical decoupling between the crossbar and the post-neuron stage.
In inference/read operation, Fig.\ref{fig4}(b)-top, the CCII-based FIB is used to establish defined voltage and load-line conditions at the crossbar-column interface while conveying the accumulated column current to the post-neuron stage. A reference voltage applied at terminal Y sets the potential at terminal X, which behaves as a low-impedance interaction node and thereby defines the read-side load-line seen by the column. Under this condition, the accumulated column current $\sigma$I enters terminal X and is conveyed to terminal Z, from where it is delivered to the post-neuron as $I_{out}$=$\sigma$I. In the language of taxonomy, inference therefore corresponds to a composite use of the CCII-based FIB, in which a VCVS-like action establishes the desired voltage condition at the column node while a CCCS-like action transports the resulting current without reimposing the original electrical constraint of the crossbar onto the receiving stage.
In training/write-back operation, Fig.\ref{fig4}(b)-bottom, the same three-terminal interface block is reused in time-multiplexed form to apply the post-neuron stimulus back to the array. Here, the requirement is not current conveyance to the neuron, but controlled application of a programming waveform to the selected crossbar location. The CCII-based FIB therefore exploits its voltage-transfer capability to apply $V_{post}(t)$ at the relevant interaction node while maintaining isolation and controllability along the write path. The same three-terminal structure thus supports both inference and training, enabling current transfer in read mode and controlled voltage application in write mode
Accordingly, the CCII-based FIB is placed at the column-side junctions $J_9$–$J_{12}$, as shown in Fig.\ref{fig4}(c).

\section{Experimental Validation of the CCII-based Interface Strategy}
\begin{figure}[H]
    \centering
        \includegraphics[width=0.9\textwidth]{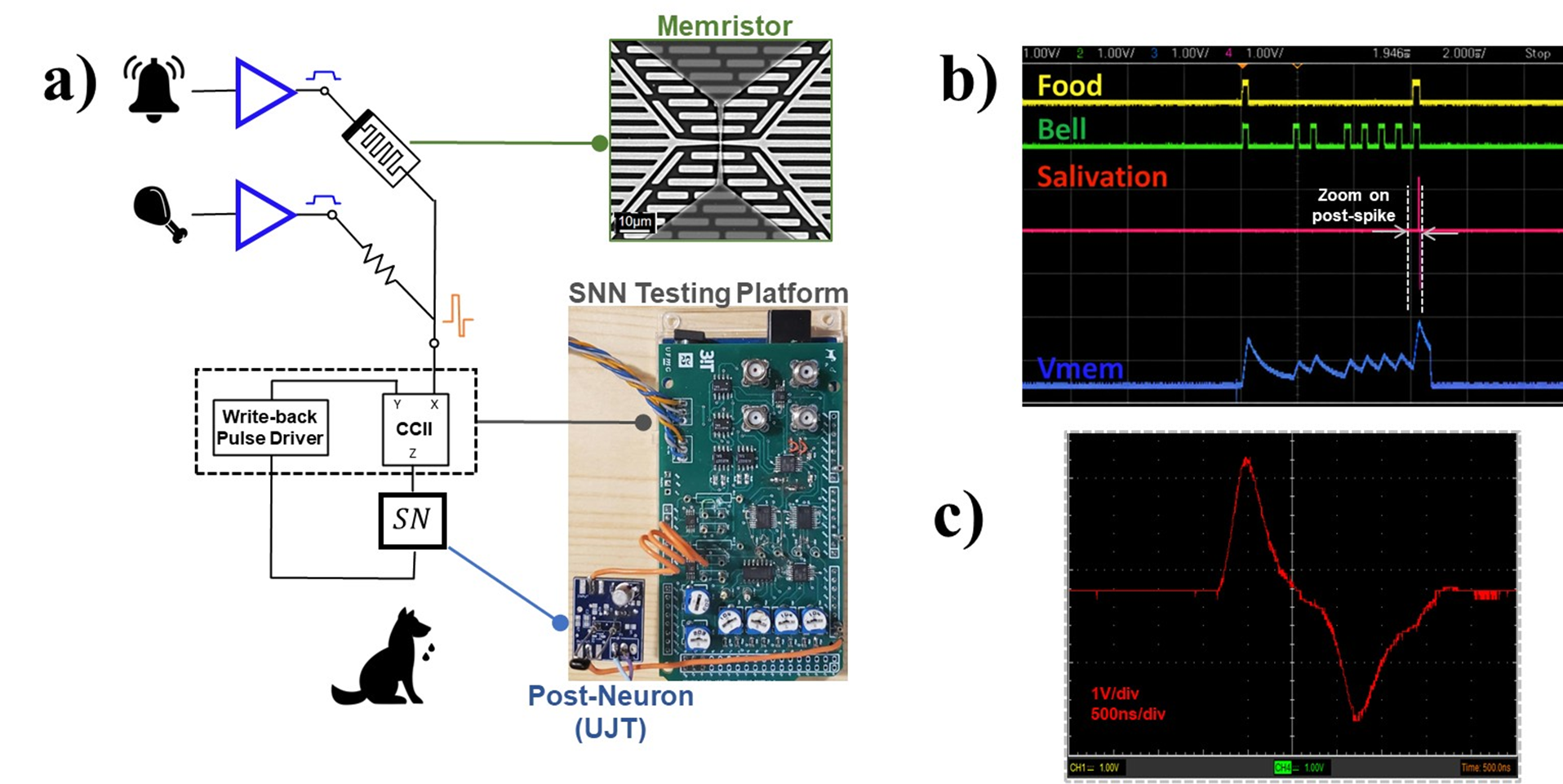}
    \caption{
        \justifying
Experimental demonstration of Pavlovian conditioning with a memristive synapse and a UJT-based post-neuron enabled by a CCII interface. (a) Simplified test setup showing the memristor synapse, the CCII-based interface (FIB), the bipolar write-back pulse driver used for plasticity, and the post-neuron implemented with a unijunction transistor (UJT). (b) Oscilloscope capture of a representative event, showing the synchronized Food, Bell, Salivation, and membrane potential Vmem traces, with an inset zoom of the post-spike waveform.}
\label{fig5}
\end{figure}

\justifying
The proposed interface strategy was experimentally validated using the learning demonstrator, structured around a Pavlovian-conditioning protocol, shown in Fig.\ref{fig5}(a). In this architecture, the Food pathway is represented by a fixed resistive synapse, reflecting its role as an unconditioned stimulus that activates the post-neuron. The Bell pathway is implemented with a memristive synapse, allowing its conductance to be modified during conditioning. The CCII-based FIB mediates the interaction between synapse and the UJT neuron, while the write-back driver applies the programming stimulus required to update the memristive state. Component-level details are provided in the Supplementary Material. A representative epoch is shown in Fig.\ref{fig5}(b), where oscilloscope traces capture the synchronized Food, Bell, Salivation, and membrane-potential waveforms during a single event. The spike amplitude, leakage resistance, and membrane capacitance were adjusted to obtain an epoch duration of approximately 8 ms. The inset in Fig.\ref{fig5}(c) details the bipolar write-back spike, whose polarity-asymmetric shape follows the STDP learning rule reported in \cite{Querlioz2013DeviceVariations}, while its nanosecond-scale pulse width is adjusted to meet the programming requirements of the memristive synapse \cite{ElMesoudy2022CMOSMemristorCrossbars}.

\begin{figure}[H]
    \centering
        \includegraphics[width=0.8\textwidth]{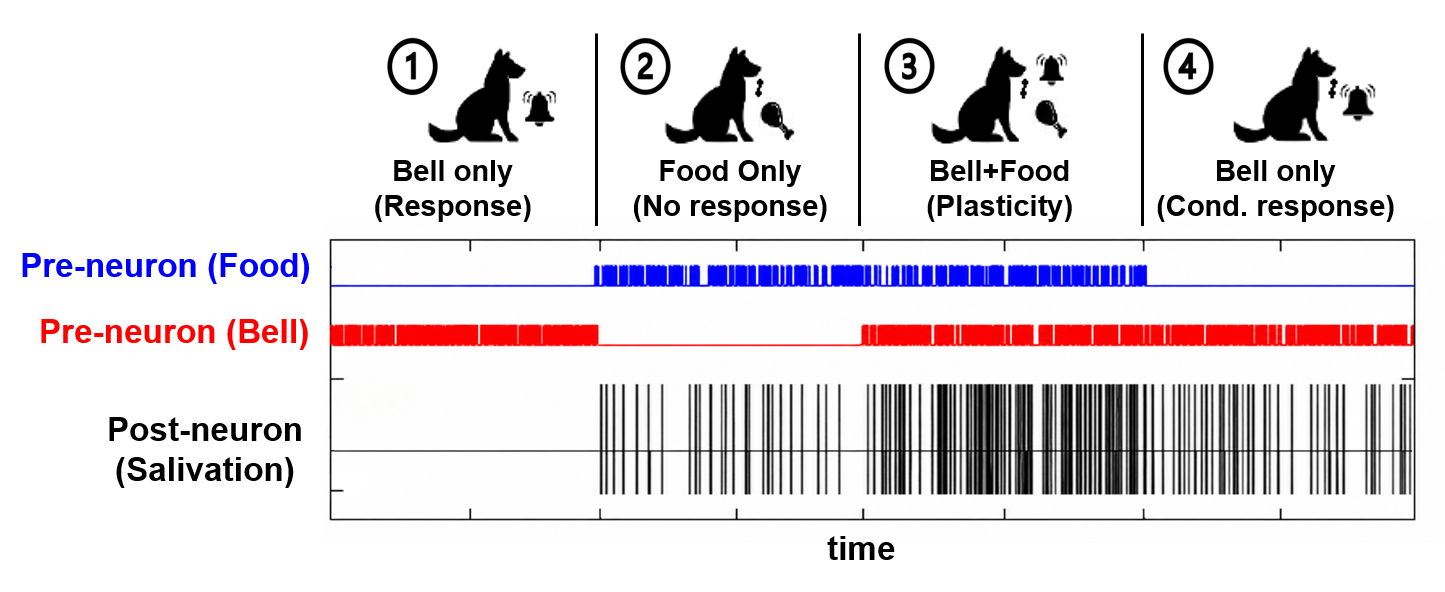}
    \caption{
        \justifying
Pavlovian-conditioning validation protocol. Experimental results showing the four-phase Pavlovian protocol illustrated by spike rasters for the pre-neurons (Food/US and Bell/CS) and the post-neuron response (Salivation).}
\label{fig6}
\end{figure}

\noindent The demonstrator was controlled by an Arduino Mega 2560, but its processing resources were not used to emulate or interfere with the SNN dynamics. Importantly, the Arduino is electrically and functionally outside the neuronal signal path: the microcontroller acted only as a supervisory unit, sequencing the conditioning phases, recording neuronal activity, pausing the network between epochs, and triggering monitoring cycles to read the synaptic resistance. Therefore, the post-neuron response and the activity-dependent synaptic adaptation were produced by the analog CCII–memristor neuromorphic hardware. The Arduino also introduced spike-time jitters in the Food and Bell drivers, preventing strictly periodic inter-spike intervals and producing more biologically plausible temporal variability. 
The validation protocol follows, summarized in Fig.\ref{fig6}, described by time-domain windows plots. The sequence begins with a Bell-only pre-conditioning phase, in which the conditioned stimulus does not yet elicit a post-neuron response. This is followed by a Food-only phase, where the unconditioned stimulus reliably triggers the output response. During the Bell-plus-Food pairing phase, the post-neuron activity and write-back pathway induce plasticity in the memristive Bell synapse. Finally, in the Bell-only post-conditioning phase, the learned association is evaluated by verifying whether the Bell stimulus alone can activate the post-neuron. Following non-volatile synaptic behavior, the Bell-only was applied showing that this behavior can be disassociated. This protocol therefore tests both functional signal propagation and activity-dependent conductance adaptation within the same heterogeneous neuromorphic architecture.

\begin{figure}[H]
    \centering
        \includegraphics[width=0.8\textwidth]{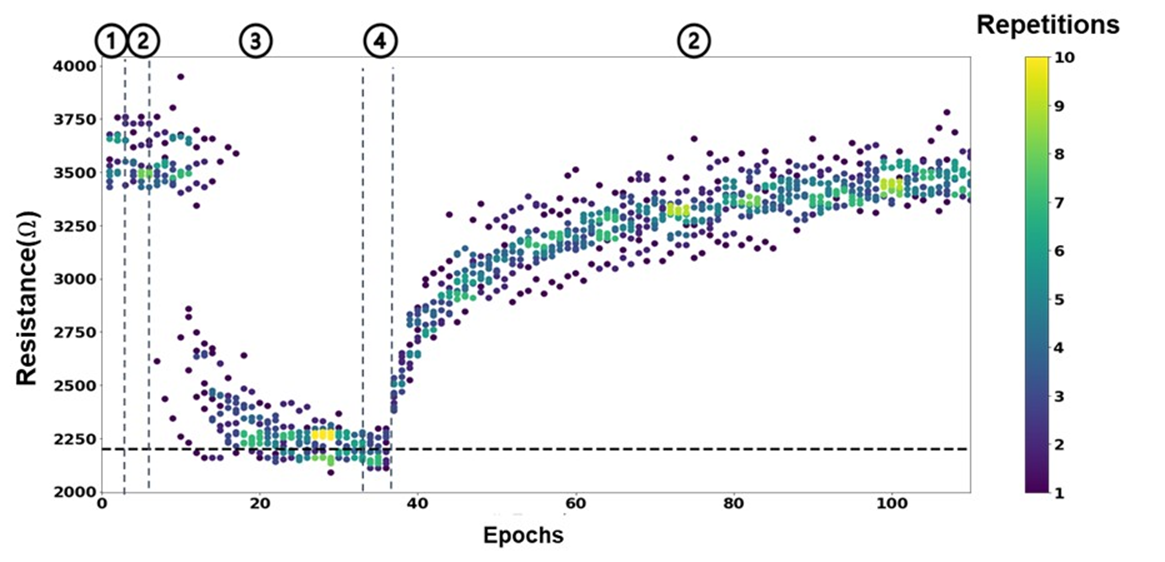}
    \caption{
        \justifying
Pavlovian-conditioning validation protocol. Experimental results showing the four-phase Pavlovian protocol illustrated by spike rasters for the pre-neurons (Food/US and Bell/CS) and the post-neuron response (Salivation).}
\label{fig7}
\end{figure}

\noindent The evolution of the memristive synapse during the training and validation protocol is summarized in Fig.\ref{fig7}, which reports the resistance trajectory across epochs as an occurrence-coded map over ten runs. The initial Bell-only and Food-only stages, corresponding to phases 1 and 2 in Fig.\ref{fig7}, establish the baseline behavior: the Bell input does not yet activate the post-neuron, whereas the Food input reliably produces the unconditioned response. During the Bell-plus-Food pairing stage, phase 3 in Fig. \ref{fig7}, the measured resistance progressively decreases, indicating conductance strengthening of the memristive Bell synapse. This change enables the transition from Bell-insensitive to Bell-responsive behavior, as confirmed in the subsequent Bell-only validation stage, phase 4 in Fig.\ref{fig7}. This behavior is consistent with the Hebbian principle \cite{Hebb1949Organization} that “neurons that fire together wire together,” since repeated temporal correlation between Bell activity and the salivation/post-neuron response strengthens the Bell synapse. The clustering of repeated resistance values in similar epoch regions indicates that this synaptic evolution is reproducible across runs despite device- and cycle-level variability.

\noindent Beyond the average trend, the occurrence-coded representation provides a statistical view of the protocol repeatability: high-occurrence regions indicate that multiple runs reached similar resistance states at corresponding epochs, whereas lower-occurrence regions capture run-to-run variability. The synaptic resistance profile shows not only the expected conductance modulation during learning, but also the resilience of the Pavlovian conditioning protocol, since the learned Bell-responsive state is consistently reached despite the intrinsic variability of the memristive synapse. A final extinction/forgetting phase was then introduced by removing the Bell–Food pairing condition, phase 2 in Fig.\ref{fig7}. Under this condition, the synaptic state gradually relaxes toward a non-conditioned regime, showing that the learned association can be weakened when correlated activity is no longer reinforced. Importantly, this forgetting behavior was intentionally induced by the protocol; once write-back pulses are disabled, the memristive synapse preserves its programmed state according to its non-volatile characteristics. Together, the temporal measurements and resistance statistics support the central experimental claim of this work: the CCII-based FIB preserves stable heterogeneous operation while enabling write-back-driven plasticity in the learning demonstrator.

\section{Conclusion}
\justifying
In this work, we show that the operation of physical neuromorphic systems is defined not solely by device physics, but by the electrical conditions established at the junctions where heterogeneous elements interact. This is a fundamental constraint in neuromorphic hardware, because synaptic readout, state programming and neuronal activation frequently require distinct, and often incompatible, biasing conditions while remaining electrically- coupled. We addressed this challenge by introducing a junction-centred framework in which interfacing is elevated to an explicit functional primitive, formalized here as the functional interface block. Framed in terms of drive-sense relations across ports, the FIB provides a general basis for analysing and engineering electrically mediated interactions in heterogeneous neuromorphic assemblies.
A CCII-based interface was used to experimentally validate the framework in an associative memristive platform combining synaptic plasticity, bipolar write-back and post-neuronal activation. The experiments demonstrate that functional decoupling at the shared junction can be achieved without suppressing the desired system-level behaviour, thereby enabling controlled stimulus association together with direct observation of synaptic state evolution. Although the present realization is constrained by the practical limits of active CCII circuitry, including finite operating range and scalability overhead, these limitations do not diminish the broader implication of the work: that successful neuromorphic integration requires the interface itself to be treated as a first-class design object.
More generally, the FIB concept points towards a route by which diverse emerging devices may be assembled into coherent physical systems, offering a framework for neuromorphic hardware that is not only functional, but also structurally interpretable and technologically extensible. An important next step will be to develop highly scalable interface implementations that retain the functional role demonstrated here by the CCII, while defining quantitative measures of interface efficiency that allow the coupling layer itself to be engineered and optimized beyond the intrinsic constraints of device physics.

\bibliographystyle{iopart-num}
\bibliography{references_fib_iop}

@article{Adhikari2012MemristorBridge,
  author  = {Adhikari, Shyam Prasad and others},
  title   = {Memristor bridge synapse-based neural network and its learning},
  journal = {IEEE Transactions on Neural Networks and Learning Systems},
  volume  = {23},
  number  = {9},
  pages   = {1426--1435},
  year    = {2012}
}

@book{AlexanderSadiku2021Fundamentals,
  author    = {Alexander, Charles K. and Sadiku, Matthew N. O.},
  title     = {Fundamentals of Electric Circuits},
  edition   = {7},
  publisher = {McGraw-Hill},
  year      = {2021}
}

@article{Brown2024AxonLike,
  author  = {Brown, Timothy D. and others},
  title   = {Axon-like active signal transmission},
  journal = {Nature},
  volume  = {633},
  number  = {8031},
  pages   = {804--810},
  year    = {2024}
}

@article{Chicca2014NeuromorphicCircuits,
  author  = {Chicca, Elisabetta and Stefanini, Fabio and Bartolozzi, Chiara and Indiveri, Giacomo},
  title   = {Neuromorphic electronic circuits for building autonomous cognitive systems},
  journal = {Proceedings of the IEEE},
  volume  = {102},
  number  = {9},
  pages   = {1367--1388},
  year    = {2014}
}

@article{Del2019subthreshold,
  title={Subthreshold firing in Mott nanodevices},
  author={Del Valle, Javier and Salev, Pavel and Tesler, Federico and Vargas, Nicol{\'a}s M and Kalcheim, Yoav and Wang, Paul and Trastoy, Juan and Lee, Min-Han and Kassabian, George and Ram{\'\i}rez, Juan Gabriel and others},
  journal={Nature},
  volume={569},
  number={7756},
  pages={388--392},
  year={2019},
  publisher={Nature Publishing Group UK London}
}

@article{ElMesoudy2022CMOSMemristorCrossbars,
  author  = {El Mesoudy, Abdelouadoud and others},
  title   = {Fully CMOS-compatible passive TiO2-based memristor crossbars for in-memory computing},
  journal = {Microelectronic Engineering},
  volume  = {255},
  pages   = {111706},
  year    = {2022}
}

@inproceedings{Garg2024AnalogLIF,
  author    = {Garg, Nikhil and others},
  title     = {Versatile CMOS analog LIF neuron for memristor-integrated neuromorphic circuits},
  booktitle = {2024 International Conference on Neuromorphic Systems (ICONS)},
  publisher = {IEEE},
  year      = {2024}
}

@article{Grollier2020NeuromorphicSpintronics,
  author  = {Grollier, Julie and others},
  title   = {Neuromorphic spintronics},
  journal = {Nature Electronics},
  volume  = {3},
  number  = {7},
  pages   = {360--370},
  year    = {2020}
}

@book{Hebb1949Organization,
  author    = {Hebb, Donald O.},
  title     = {The Organization of Behavior: A Neuropsychological Theory},
  publisher = {Wiley},
  year      = {1949}
}

@article{Hodgkin1952propagation,
  title={Propagation of electrical signals along giant nerve fibres},
  author={Hodgkin, Alan Lloyd and Huxley, Andrew Fielding},
  journal={Proceedings of the Royal Society of London. Series B-Biological Sciences},
  volume={140},
  number={899},
  pages={177--183},
  year={1952},
  publisher={The Royal Society London}
}

@article{Huang2024MemristorAccelerators,
  author  = {Huang, Yi and others},
  title   = {Memristor-based hardware accelerators for artificial intelligence},
  journal = {Nature Reviews Electrical Engineering},
  volume  = {1},
  number  = {5},
  pages   = {286--299},
  year    = {2024}
}

@article{Kim2024BiomimeticSpikingNeuron,
  author  = {Kim, Yijoon and others},
  title   = {Highly biomimetic spiking neuron using SiGe heterojunction bipolar transistors for energy-efficient neuromorphic systems},
  journal = {Scientific Reports},
  volume  = {14},
  number  = {1},
  pages   = {8356},
  year    = {2024}
}

@article{Koo2025CommercialEmergingMemory,
  author  = {Koo, Ryun-Han and others},
  title   = {From commercial to emerging memory: Device and circuit perspectives for neuromorphic computing},
  journal = {Device},
  year    = {2025}
}

@article{Indiveri2011SiliconNeuronCircuits,
  author  = {Indiveri, Giacomo and others},
  title   = {Neuromorphic silicon neuron circuits},
  journal = {Frontiers in Neuroscience},
  volume  = {5},
  pages   = {73},
  year    = {2011}
}

@inproceedings{Lecerf2014silicon,
  author    = {Lecerf, Gwendal and Tomas, Jean and Boyn, S{\"o}ren and Girod, St{\'e}phanie and Mangalore, Ashwin and Grollier, Julie and Sa{\"i}ghi, Sylvain},
  title     = {Silicon Neuron Dedicated to Memristive Spiking Neural Networks},
  booktitle = {2014 IEEE International Symposium on Circuits and Systems (ISCAS)},
  pages     = {1568--1571},
  year      = {2014},
  address   = {Melbourne, VIC, Australia},
  publisher = {IEEE},
  doi       = {10.1109/ISCAS.2014.6865448}
}

@article{Li2023MemristiveNeuronSynapse,
  author  = {Li, Jiayi and others},
  title   = {Emerging memristive artificial neuron and synapse devices for the neuromorphic electronics era},
  journal = {Nanoscale Horizons},
  volume  = {8},
  number  = {11},
  pages   = {1456--1484},
  year    = {2023}
}

@article{Mahowald1991silicon,
  title={A silicon neuron},
  author={Mahowald, Misha and Douglas, Rodney},
  journal={Nature},
  volume={354},
  number={6354},
  pages={515--518},
  year={1991},
  publisher={Nature Publishing Group UK London}
}

@article{Markovic2020physics,
  title = {Physics for neuromorphic computing},
  author = {Markovi{\'c}, Danijela and Mizrahi, Alice and Querlioz, Damien and Grollier, Julie},
  journal = {Nature Reviews Physics},
  volume = {2},
  number = {9},
  pages = {499--510},
  year = {2020},
  publisher = {Nature Publishing Group UK London}
}

@article{Mead2002NeuromorphicElectronicSystems,
  author  = {Mead, Carver},
  title   = {Neuromorphic electronic systems},
  journal = {Proceedings of the IEEE},
  volume  = {78},
  number  = {10},
  pages   = {1629--1636},
  year    = {2002}
}

@article{Musil2009CurrentConveyor,
  author  = {Musil, Roman and Prokop, Vladislav and Prokop, R.},
  title   = {Current Conveyor CCII as the Most Versatile Analog Circuit Building Block},
  journal = {Annual Journal of Electronics},
  pages   = {73--76},
  year    = {2009}
}

@article{Pickett2013ScalableNeuristor,
  author  = {Pickett, Matthew D. and Medeiros-Ribeiro, Gilberto and Williams, R. Stanley},
  title   = {A scalable neuristor built with Mott memristors},
  journal = {Nature Materials},
  volume  = {12},
  number  = {2},
  pages   = {114--117},
  year    = {2013}
}

@article{Querlioz2013DeviceVariations,
  author  = {Querlioz, Damien and others},
  title   = {Immunity to device variations in a spiking neural network with memristive nanodevices},
  journal = {IEEE Transactions on Nanotechnology},
  volume  = {12},
  number  = {3},
  pages   = {288--295},
  year    = {2013}
}

@article{Rozenberg2019UltraCompactLIF,
  author  = {Rozenberg, M. J. and Schneegans, O. and Stoliar, P.},
  title   = {An ultra-compact leaky-integrate-and-fire model for building spiking neural networks},
  journal = {Scientific Reports},
  volume  = {9},
  number  = {1},
  pages   = {11123},
  year    = {2019}
}

@article{Sarkar2022OrganicSpikingNeuron,
  author  = {Sarkar, Tanmoy and others},
  title   = {An organic artificial spiking neuron for in situ neuromorphic sensing and biointerfacing},
  journal = {Nature Electronics},
  volume  = {5},
  number  = {11},
  pages   = {774--783},
  year    = {2022}
}

@article{SedraSmith1970CurrentConveyor,
  author  = {Sedra, Adel and Smith, Kenneth},
  title   = {A second-generation current conveyor and its applications},
  journal = {IEEE Transactions on Circuit Theory},
  volume  = {17},
  number  = {1},
  pages   = {132--134},
  year    = {1970}
}

@book{Sedra2011MicroelectronicCircuits,
  author    = {Sedra, Adel S. and others},
  title     = {Microelectronic Circuits},
  volume    = {4},
  publisher = {Oxford University Press},
  address   = {Oxford},
  year      = {2011}
}

@article{Sowmya2023NeuromorphicFPGA,
  author  = {Sowmya, Nagavarapu and others},
  title   = {Neuromorphic processor design and FPGA implementation for handwritten digits employing spiking neural network},
  journal = {International Journal of Computing and Digital Systems},
  volume  = {14},
  number  = {1},
  pages   = {1--1},
  year    = {2023}
}

@article{Wang2015TaTaOxTiO2SynapticDevice,
  author  = {Wang, Y.-F. and others},
  title   = {Characterization and Modeling of Nonfilamentary Ta/TaO$_x$/TiO$_2$/Ti Analog Synaptic Device},
  journal = {Scientific Reports},
  volume  = {5},
  pages   = {10150},
  year    = {2015}
}

@article{Wang2020ResistiveSwitchingMaterials,
  author  = {Wang, Zhongrui and others},
  title   = {Resistive switching materials for information processing},
  journal = {Nature Reviews Materials},
  volume  = {5},
  number  = {3},
  pages   = {173--195},
  year    = {2020}
}

@article{Yang2008memristive,
  title={Memristive switching mechanism for metal/oxide/metal nanodevices},
  author={Yang, J Joshua and Pickett, Matthew D and Li, Xuema and Ohlberg, Douglas AA and Stewart, Duncan R and Williams, R Stanley},
  journal={Nature nanotechnology},
  volume={3},
  number={7},
  pages={429--433},
  year={2008},
  publisher={Nature Publishing Group UK London}
}

@article{ZamarrenoRamos2011STDP,
  author  = {Zamarre{\~n}o-Ramos, Carlos and others},
  title   = {On spike-timing-dependent-plasticity, memristive devices, and building a self-learning visual cortex},
  journal = {Frontiers in Neuroscience},
  volume  = {5},
  pages   = {26},
  year    = {2011}
}

\ack{The authors acknowledge the support of 3iT (Interdisciplinary Institute for Technological Innovation) and the Department of Physics of the Universidade Federal de Minas Gerais (UFMG). The authors also thank the colleagues and collaborators who contributed through technical discussions on neuromorphic hardware, memristive devices, and circuit-level interfacing during the development of this work. }

\section*{Conflict of interest}
The authors declare no competing interests.

\section*{Ethical statement}
This study did not involve human participants, human data, human tissue, or animals. Therefore, ethics approval was not required.

\funding{This work was supported in part by the Conselho Nacional de Desenvolvimento Científico e Tecnológico (CNPq) and the Fundação de Amparo à Pesquisa do Estado de Minas Gerais (FAPEMIG). W. A. acknowledges support from Mitacs through the Mitacs Globalink Research Award.}

\roles{WA: Conceptualization, Methodology, Investigation, Resources, Data curation, Formal analysis, Visualization, Writing – original draft.

\noindent GMR: Conceptualization, Methodology, Formal analysis, Supervision, Writing – review \& editing.

\noindent YB: Validation, Writing – review \& editing.

\noindent FA: Validation, Writing – review \& editing.

\noindent DD: Validation, Writing – review \& editing.

\noindent All authors reviewed and approved the final manuscript.}

\data{The data supporting the findings of this study are included within the paper and its supplementary material. The numerical data underlying the plotted experimental results are provided as Supplementary Data. No simulation code or custom analysis software was generated for this study.}


\suppdata{Supplementary material is available online. It includes additional methodological details, supporting experimental records, and extended information related to the hardware implementation and interpretation of the results.}

\end{document}